\documentclass[aps,twocolumn,amsmath,amssymb,floatfix]{revtex4}
\usepackage{graphicx}
\usepackage{epstopdf}
\begin{document}


\markboth{Kia \textit{et al.}}{Coupling Reduces Noise}

\title{COUPLING REDUCES NOISE}

\author{BEHNAM KIA$^1$, SARVENAZ KIA$^1$, JOHN F. LINDNER$^2$, \\SUDESHNA SINHA$^3$, WILLIAM L. DITTO$^1$}

\address{$^1$Department of Physics and Astronomy, University of Hawai`i at M\=anoa \\
Honolulu, Hawai`i 96822, USA}

\address{$^2$Physics Department, The College of Wooster
\\Wooster, Ohio 44691, USA}

\address{$^3$Indian Institute of Science Education and Research (IISER)\\
Mohali, Punjab, India}

\begin{abstract}
We demonstrate how coupling nonlinear dynamical systems can reduce the effects of noise. For simplicity we investigate noisy coupled map lattices. Noise from different lattice nodes can diffuse across the lattice and lower the noise level of individual nodes. We develop a theoretical model that explains this observed noise evolution and show how the coupled dynamics can naturally function as an averaging filter. Our numerical simulations are in excellent agreement with the model predictions. 
\end{abstract}

\maketitle




\section{Introduction}

We investigate the evolution of local noise under coupled dynamics and demonstrate how coupled dynamics reduces the overall noise content of the individual elements. For simplicity we consider a coupled map lattice (CML) as a paradigm for generically coupled systems \cite{Kaneko}. We model how coupling identical maps can reduce the noise content of the overall lattice and observe that the noise from different nodes diffuses across the lattice and attenuates the noise in each element of the lattice. 

Additionally we demonstrate that an implicit form of average filtering, also known as ensemble averaging, can occur within the coupled map dynamics. An averaging filter is a widely used noise filtering technique where different readings of a measurement, or different values of local independent nodes in a spatially extended system, are averaged and the result is a system with less noise content \cite{Smith}. In this work we demonstrate how one can tune the coupled dynamics to act as an averaging filter for noise in locally and globally coupled systems This dynamics-based realization of an averaging filter has practical importance in applications that utilize dynamics to operate. Such applications include optimizers for hard problems \cite{Ercsey}, chaos based data transmission \cite{Grebogi}, and chaos based computation \cite{Sinha}. Evaluating and improving robustness to noise in such dynamics based applications is of critical importance. Extensive research has been dedicated to investigate and improve the robustness to noise in such systems \cite{Sumi, Dolnik, Kia}. From the practical standpoint our observation and development of coupled dynamics as a noise filtering mechanism is novel because it allows us to utilize the inherent coupled dynamics for noise reduction rather than resorting to an external filter or different design that applies an averaging type filter for noise reduction. Rather than having one isolated, independent dynamical system to implement the application, we can use a set of similar dynamical systems, which are dynamically coupled together, to implement noise reduction. This has particular application to chaotic computation where noise can play either a catastrophic or productive role in computation \cite{Kia, Murali}.

This noise reduction mechanism can filter noise from the steady-state, long term behavior of periodic CMLs as well as from its transient behavior. In chaotic CMLs, this technique can only reduce noise from the transient behavior of the CML. However, many dynamics based applications such as chaotic computation, chaotic optimization, or chaotic data transmission utilize this transient behavior. As a result, this noise filtering of transient chaos is of particular relevance for such applications. 

It is well known that when signal and noise are added to each node of a coupled array, the signal adds coherently but the noise adds incoherently \cite{Gammaitoni}, thus increasing the signal-to-noise ratio. Furthermore arranging bistable dynamical systems in an array and coupling them together enhances their response a time-periodic signal by increasing a signal-to-noise ratio \cite{Lindner}. Here we focus on the ability of dynamical coupling to reduce the effects noise.

The organization of the paper is as follows. In Section~\ref{CMLSection} we describe the coupled map model that we use. In Section~\ref{NoiseTheorySection} we derive a model for noise deviation in the coupled map lattice, which is the deviation from noise free evolution caused by noise, and using this model we explain and predict the mechanism through which the dynamics of the coupled systems reduces the noise. In Section~\ref{SimSection} we present our simulation results, and in Section~\ref{Conclusions} we discuss our conclusions.

\section{Coupled Map Lattice}	\label{CMLSection}

To demonstrate noise reduction by coupling, we first study coupled map lattices (CMLs). The model for noise deviation is general and applicable to a variety of dynamics. The CML we use to first demonstrate noise reduction was introduced by Kaneko \cite{Kaneko} and has been extensively studied. It combines a nonlinear map with diffusive coupling in the recurrence relation
\begin{equation} \label{KanekoEq}
	x_n^{i+1} 
	= (1 - \epsilon) f[ x_n^i ] + \frac{\epsilon}{2} \left( f[x_{n-1}^i] + f[x_{n+1}^i] \right),
\end{equation}
where $x_n^i$ is the state of the $n$th node at the $i$th iteration, $n = 1, 2, 3, \ldots, N$ and $i = 1, 2, 3, \ldots$, $f[x]$ is a one-dimensional map, and $\epsilon$ is the coupling parameter. The first and the last map of the lattice are connected to each other to realize periodic boundary conditions. For simplicity we start with a small lattice of size $N=3$, where local and global coupling are the same, and later generalize to a globally coupled map lattice of any size. 

\section{Theoretical Analysis} \label{NoiseTheorySection}

We model the effects of noise and coupling on the Eq.~(\ref{KanekoEq}) CML by comparing the deviation of the noisy and noiseless evolutions of the coupled system to the deviation of the noisy and noiseless evolution of the uncoupled single isolated map. To broadly quantify this effect, we define noise robustness 
\begin{equation} \label{REq}
	R 
	= \frac{\sigma_s^2}{\sigma_c^2},
\end{equation}
where $\sigma_s^2$  is the variance of noise deviation in a single isolated map and $\sigma_c^2$ is the variance of noise deviation in $N$ coupled maps. The maps are subjected to local, statistically independent but identically distributed noise. All the CML maps have a noisy version of the same initial condition $x^0$ of the single isolated map.

Specifically, the initial conditions of the maps in the CML are
\begin{equation}
	x_n^0 
	= x^0 + \sigma \delta_n^{0},
\end{equation}
and the noise is added to each node at each future iteration like
\begin{equation} \label{iterateEq}
	x_n^{i+1} 
	= f \left[ x_n^i \right] + \sigma \delta_n^{i+1},
\end{equation}
where $\delta_n^i$ is normal Gaussian noise with zero mean and unit variance. The variance $\sigma^2$ of the maps' additive noises is the same, and the noise terms are independent and identically distributed for different maps of the CML and for different iterations of the maps. 

The future noisy state of a map in the CML after one iteration or step will be
\begin{align} 
	x_n^1 &= (1 - \epsilon) f \left[ x_n^0  \right]  \nonumber \\
	&+ \frac{\epsilon}{2} \left( f \left[ x_{n-1}^0 \right] + f \left[ x_{n+1}^0 \right] \right) + \sigma \delta_n^1 \nonumber \\
	&= (1 - \epsilon) f \left[  x^0 + \sigma \delta_n^0  \right]  \nonumber\\
	&+ \frac{\epsilon}{2} \left( f \left[  x^0 + \sigma \delta_{n-1}^0  \right] + f \left[  x^0 + \sigma \delta_{n+1}^0  \right] \right) + \sigma \delta_n^1,
\end{align}
which, after linearizing the function $f[x]$ around the initial condition $x^0$, becomes
\begin{align} \label{LinearizedEq}
	x_n^1 
	&= f[x^0] \nonumber \\
	&+ \sigma \lambda_1 \left( (1 - \epsilon) \delta_n^0 + \frac{\epsilon}{2} \delta_{n-1}^0 + \frac{\epsilon}{2} \delta_{n+1}^0 \right) + \sigma \delta_n^1,
\end{align}
where $\lambda_1 = df / dx$  is evaluated at $x^0$. The one-step map is composed of three elements: (1) noise free evolution of the map starting from the noise free initial condition, $f[ x_0 ]$; (2) evolution of previous noise terms under the coupled dynamics, $\sigma \lambda_1 \left( (1 - \epsilon) \delta_n^0 + (\epsilon/2) \delta_{n-1}^0 + (\epsilon/2) \delta_{n+1}^0 \right)$; and (3) noise term that is added to the map at the current iteration, $\sigma \delta_n^1$. The term of interest is the second, which models the noise deviation in the coupled dynamics. The noise terms added to different maps are independent and identically distributed. Because independent uncertainties add in quadrature, the variance of the second term is
\begin{align} \label{oneStepEq}
	\sigma_{c1}^2 
	&= \sigma^2 \lambda_1^2 \left( (1-\epsilon)^2 \sigma_\delta^2 + \frac{\epsilon^2}{4} \sigma_\delta^2 + \frac{\epsilon^2}{4} \sigma_\delta^2 \right) \nonumber \\
	& = \left( \sigma \lambda_1 \right)^2 \left( (1 - \epsilon)^2 + \frac{\epsilon^2}{2} \right),
\end{align}
which is the coupled map one-step noise variance. Setting $\epsilon = 0$ in Eq.~\ref{oneStepEq} gives
\begin{equation} \label{epsilonREq}
	\sigma_{s1}^2 
	= \left( \sigma \lambda_1 \right)^2
	= \left( (1 - \epsilon)^2 + \frac{\epsilon^2}{2} \right)^{-1} \sigma_{c1}^2,
\end{equation}
which is the single isolated map one-step noise variance. 

Comparing Eq. (\ref{epsilonREq}) and Eq. (\ref{REq}), the CML one-step noise robustness $\smash{ R = \left( (1 - \epsilon)^2 + \epsilon^2/2 \right)^{-1} }$. For coupling $0 < \epsilon < 4/3$, the noise robustness $R > 1$, and the variance of noise deviation in the CML is less than the variance of noise deviation in a single isolated map. Furthermore, an optimal coupling parameter of $\epsilon = 2/3$ implies
\begin{equation} 
	\sigma_{s1}^2 
	= \left( \sigma \lambda_1 \right)^2
	= 3 \sigma_{c1}^2
\end{equation}
and a maximum one-step noise robustness of $R = 3$. Thus, in a CML of size $N = 3$ with optimal coupling, the variance of one-step noise deviation is reduced to 1/3 of the variance when the maps are isolated. 

The same modeling and the same result can be obtained for further iterations of the CML and variance of noise deviation. Iterating Eq.~(\ref{iterateEq}) one more time with $\epsilon = 2/3$  gives
\begin{align} 
	x_n^2 
	&= \frac{1}{3} f\left[ f\left[ x^0 \right] + \sigma \lambda_1 \frac{ \delta_{n-1}^0 + \delta_n^0 + \delta_{n+1}^0}{3} + \sigma \delta_n^1 \right] \nonumber \\
	&+ \frac{1}{3} f\left[ f\left[ x^0 \right] + \sigma \lambda_1 \frac{ \delta_{n-1}^0 + \delta_n^0 + \delta_{n+1}^0 }{3} + \sigma \delta_{n-1}^1 \right]  \nonumber \\
	&+ \frac{1}{3} f\left[ f\left[ x^0 \right] + \sigma \lambda_1 \frac{ \delta_{n-1}^0 + \delta_n^0 + \delta_{n+1}^0 }{3} + \sigma \delta_{n+1}^1 \right] \nonumber \\
	&+ \sigma \delta_n^2.
\end{align}
By linearizing this around $f \left[ x^0 \right]$, we obtain
\begin{align} 
	x_n^2 
	&= \frac{1}{3} f\left[ f\left[ x^0 \right]  \right] \nonumber \\
	&+ \lambda_2 \, \sigma \lambda_1 \frac{ \delta_{n-1}^0 + \delta_n^0 + \delta_{n+1}^0}{3}  \nonumber \\
	&+ \sigma \lambda_2  \frac{ \delta_{n-1}^1 + \delta_n^1 + \delta_{n+1}^1 }{3}  \nonumber \\
	&+ \sigma \delta_n^2,
\end{align}
where $\lambda_2 = df /dx$ is evaluated at $f \left[ x_0 \right]$.

Again we observe that the noise terms $\delta_{n-1}^1$, $\delta_n^1$, $\delta_{n+1}^1$ added during the previous step are averaged by the coupled map dynamics. Once more adding independent uncertainties in quadrature, the two-step variance of noise deviations is
\begin{equation} 
	\sigma_{c2}^2 
	= \frac{1}{3} \sigma^2 \left( \left(\lambda_1 \lambda_2 \right)^2 + \lambda_2^2 \right),
\end{equation}
whereas the two-step variance of noise deviations in a single map is 
\begin{equation} 
	\sigma_{s2}^2 
	= \sigma^2 \left( \left(\lambda_1 \lambda_2 \right)^2 + \lambda_2^2 \right)
	= 3 \sigma_{c2}^2,
\end{equation}
for a two-step noise robustness of $R = 3$.

This noise averaging by the coupled dynamics is repeated over the next iterations of the map as well. In general, the variance of noise deviation in the coupled-map lattice after $i$ iterations is
\begin{align} \label{Old12Eq}
	\sigma_{ci}^2 &= 
	\frac{1}{3} \sigma^2 \left( \left( \lambda_1 \lambda_2 \cdots \lambda_{i-1} \lambda_i \right)^2 \right. \nonumber\\	
	&\left. +  \left( \lambda_2 \lambda_3 \ldots \lambda_{i-1} \lambda_i \right)^2 + \cdots + \lambda_i^2 \right),
\end{align}
where $\lambda_i = df / dx$  is evaluated at $\smash{ f^{(i-1)} \left[ x^0 \right] }$, where $\smash{ f^{(i)}\left[ x \right] }$ means $i$ iterations of function $f[x]$, whereas the variance of noise effects in the single map is
\begin{align} \label{Old13Eq}
	\sigma_{si}^2 &= 
	\sigma^2 \left( \left( \lambda_1 \lambda_2 \cdots \lambda_{i-1} \lambda_i \right)^2 \right. \nonumber\\	
	&\left. +  \left( \lambda_2 \lambda_3 \ldots \lambda_{i-1} \lambda_i \right)^2 + \cdots + \lambda_i^2 \right) 
	= 3 \sigma_{ci}^2.
\end{align}
Thus after $i$ iterations, our linearization approximation still indicates a noise robustness $R = 3$, and the Eq.~(\ref{KanekoEq}) CML reduces the variance of noise deviation to 1/3 of the variance of noise deviation of a single map. 

\begin{figure}[htb] 
	\includegraphics[width=1\linewidth]{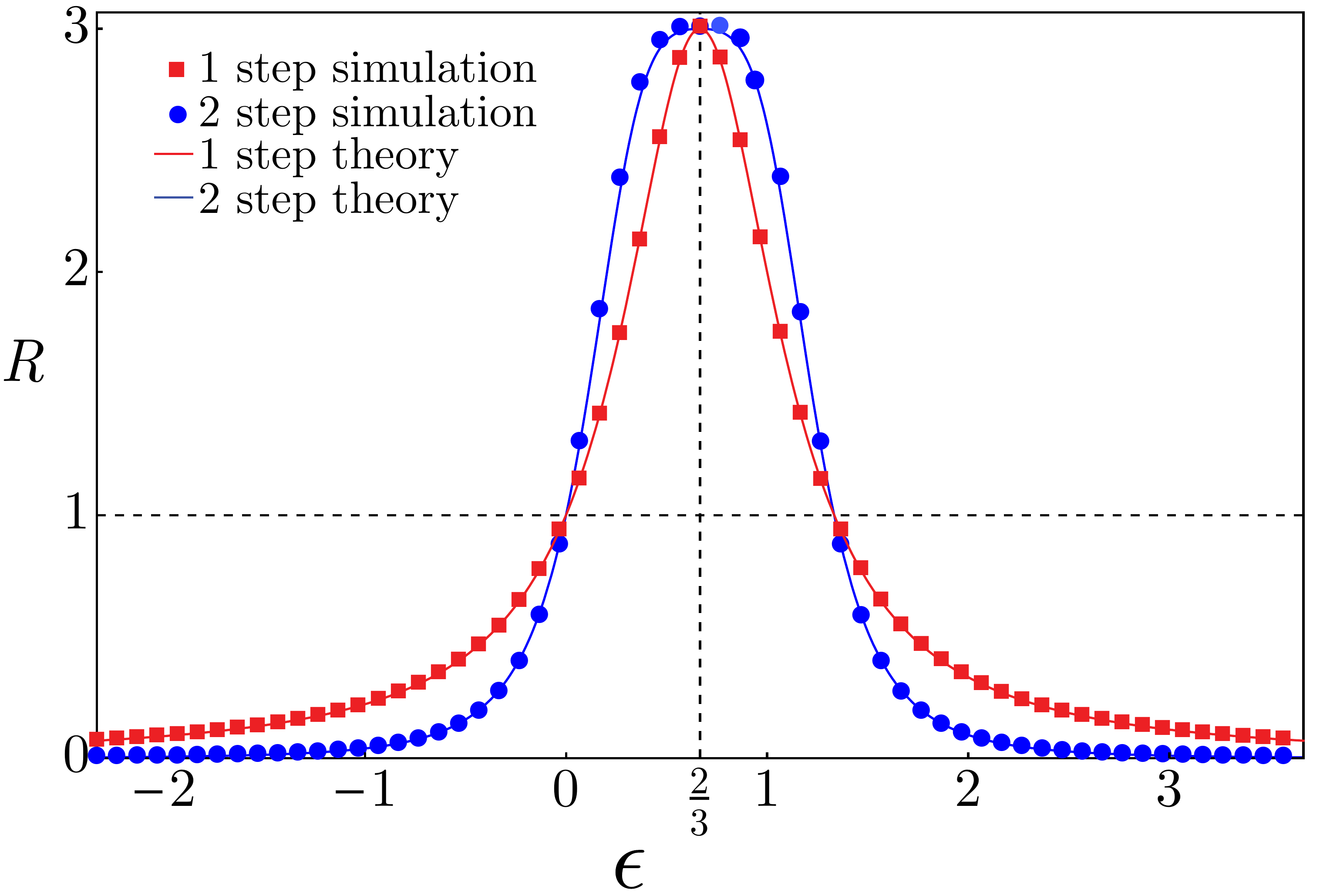}
	\caption{Noise robustness $R$ versus coupling $\epsilon$ for a ring of $N = 3$ cyclically coupled chaotic maps. Red squares indicate first iteration or step, blue disks indicate second step, and the corresponding lines indicate theory. ÊThe simulation uses the Eq. (\ref{QuadMapEq}) quadratic map and $10^6$ realizations of noise with variance $\sigma^2 = 10^{-6}$.}
	\label{RobustnessPlot}
\end{figure}

Figure~\ref{RobustnessPlot} plots noise robustness $R$ for a CML of size $N = 3$ versus different possible values of coupling parameter $\epsilon$ for one step and two step evolution of the CML. The solid lines are the theory, and data points are simulation results (which will be discussed and explained in Section~\ref{SimSection}). The peak of the noise robustness $R$ is located at $\epsilon = 2/3$ and its maximum value is $N = 3$. 
      
We can obtain this rescaling rule in a more general form for the Kaneko globally coupled map (GCM) lattice defined by
\begin{equation} \label{GCMEq}
	x_n^{i+1} 
	= (1 - \epsilon) f \left[ x_n^i \right] + \frac{\epsilon}{N-1} \sum_{m \neq n} f \left[ x_m^i \right].
\end{equation}
Similar to our analysis of 3-node CML, by linearizing GCM lattice dynamics along the noise free orbit, it is straightforward to show that the coupled dynamics of GCM shrinks the variance of noise deviation. More specifically, the optimal coupling $\epsilon = (N - 1) / N$ provides the maximum noise robustness of $R = N$, thereby reducing the variance of noise deviation of the CML to $1/N$ of the variance of noise deviation in a single map. In fact, this coupling converts the Eq.~(\ref{GCMEq}) Kaneko GCM into a filter that averages the different noise terms across the lattice,
\begin{align} 
	x_n^{i+1} 
	&= \left( 1 - \frac{N-1}{N} \right) f \left[ x_n^i \right] \nonumber \\
	&+ \frac{1/(N-1)}{N-1} \sum_{m \neq n} f \left[ x_m^i \right] \nonumber \\
	&= \frac{1}{N} \sum_{m=1}^N f \left[ x_m^i \right]
	= \left \langle f \left[ x_m^i \right] \right \rangle,
\end{align}
for all nodes $n$. In the presence of independent but identically distributed noise, the noise deviation in the coupled map is $N$ times smaller than that of an isolated map, because the variance of the mean of $N$ independent but identically distributed random variables is $1/N$ times the variance of one single random variable with the same distribution, $\sigma_c^2 = \sigma_s^2 / N$. This is a dramatic example of a GCM as the dynamical realization of an averaging filter. However, the phenomenon of coupling-reduced-noise is more general than mere averaging and exists for a broad range of coupling constants and coupling schemes.

For exact analysis and measuring of the effects of coupled dynamics on noise deviation we have excluded the noise terms that are added to the maps at the last iteration of the maps from our study. Coupled dynamics diffuses and averages the noise terms that are added from the previous steps, but it does not alter or change the effects of current local noise terms. That is why we exclude the current noise terms from our study in order to measure the effects of the coupling. For example, if we wanted to include the current noise terms, we would add $\left( \sigma \delta_n^i \right)^2$ to Eq.~(\ref{Old12Eq}) and $\left( \sigma \delta^i \right)^2$ to Eq.~(\ref{Old13Eq}). Both noise terms are independent, but identically distributed. Therefore, by including the last noise term, the ratio obtained from the analysis will not be exactly 3; rather it will depend on the last noise term. However, if the dynamics is unstable, meaning that the noise effects grow under the dynamics, we can safely assume that the amplitude of additive noise at the last step is very small compared to the evolved noise effects under the nonlinear dynamics, and therefore we can neglect the last noise term. 

Linearization of the coupled dynamics is the basis for our analytical results, and we have limited our study to a first order approximation. In the Taylor expansion of the coupled dynamics, the first order terms are the local noise terms multiplied by the local eigenvalues. These different local noise terms are summed together, and the result is that the first order approximation of the coupled map dynamics functions as an averaging filter. However, after some iteration, the second order terms become larger and larger. The local noise terms in the second order term of the Taylor expansion of the coupled map dynamics are squared, therefore there is no noise filtering for second order noise effects.

\section{Numerical Analysis} \label{SimSection}

Here we present two numerical examples of coupling reducing noise involving nonlinear maps with finite or infinite basins of attraction to test the limits of our theory.

\subsection{Quadratic Map} \label{QuadMapSubSection}

In this numerical example, we assume the famous quadratic map
\begin{equation} \label{QuadMapEq}
	x^{i + 1} = 1 - \lambda \left( x^{i} \right)^2,
\end{equation}
which is chaotic on the interval $x^i \in [-1,1]$ for bifurcation parameter $\lambda = 2$. First we arrange three of these maps in a lattice with a ring architecture, and couple them based on the Eq.~(\ref{KanekoEq}) coupling scheme. Then we initialize all three nodes of the lattice to a common initial condition $x^0 = -0.9$ and evolve the coupled map lattice in the presence of local additive noise. To measure the variance of the noise deviations, we stop the evolution and measure how much the final state of a map has deviated from noise free evolution, and then we repeat this process $10^6$ times, each time with the same initial condition. It doesnÕt matter from which map of the CML we choose to measure the distribution of noise deviations. All maps in the CML are symmetric and all have the same distribution of noise deviations. We repeat the same process for a single map, starting from the same initial condition, and measure the noise variance in the single map as well. The ratio of these variances is the Eq.~(\ref{REq}) noise robustness $R$, as shown in Fig.~\ref{RobustnessPlot} for one step and two steps of the CML. The simulation results follow the Section~\ref{CMLSection} theory, with peak noise robustness $R = 3$ and the optimal coupling parameter is $\epsilon = 2/3$. In this simulation, the variance of the additive Gaussian noise is $\sigma^2 = 10^{-6}$. However, similar results can be achieved by using other noise variances. 

\begin{figure}[htb] 
	\includegraphics[width=1\linewidth]{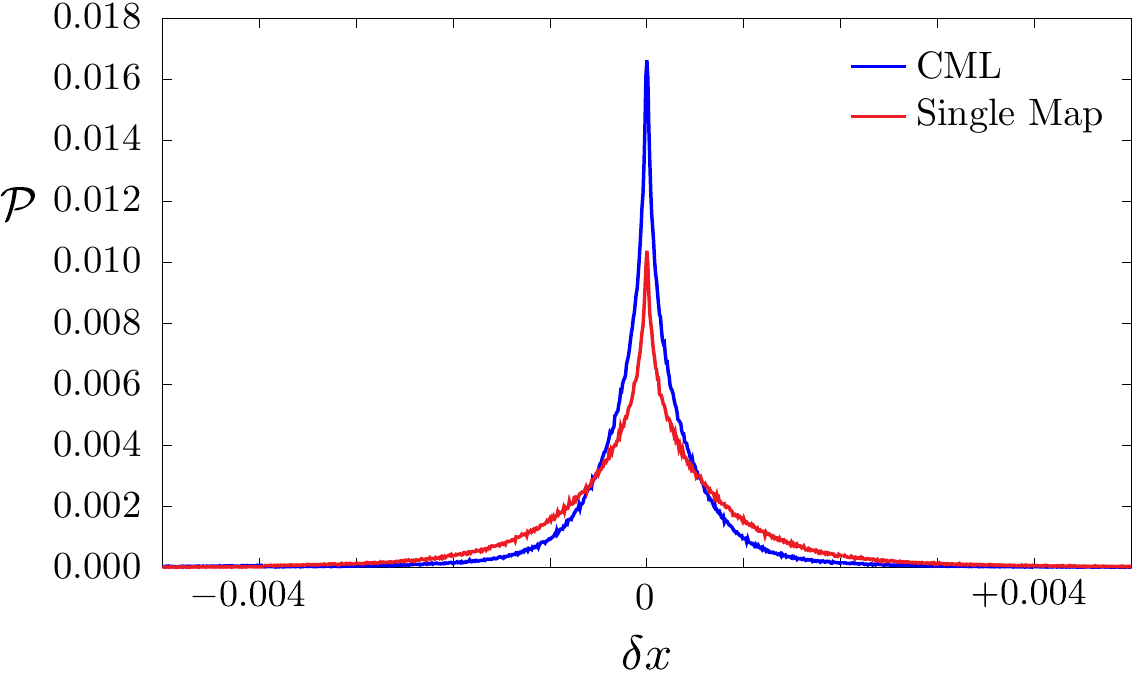}
	\caption{The probability distributions $\mathcal{P}$ of noise deviations $\delta x$ for a single isolated map (red) and for a CML of size $N=3$ (blue). The smaller spread of the CML distribution reflects the coupled dynamics suppressing the noise.}
	\label{ProbabilityDistributionPlot}
\end{figure}

The reduction of noise deviation in a CML always occurs regardless of initial condition. To further investigate this occurrence we randomly initialize all three maps of the lattice to a common point $x^0 \in [-1,1]$, which is the interval where the chaotic attractor exists. From this random initial condition we evolve the coupled map lattice for some number of iterations, and then stop the evolution and measure how much the final state of a node has deviated. The exact iteration number we used here is $i = 6$, however similar results can be obtained for other iteration numbers. Without loss of generality, we set the additive noise variance in this example to $\sigma^2 =10^{-10}$. We repeat this process $10^6$ times, each time with a new random initial condition. The probability distribution of these noise deviations is calculated and presented in Fig.~\ref{ProbabilityDistributionPlot}. We also carry out the same experiment on a single Eq.~(\ref{QuadMapEq}) map for $\lambda = 2$ . The variance of noise deviations in the coupled map lattice $\sigma_c^2 = 1.44 \times10^{-6}$, whereas the variance of noise deviations in the single map $\sigma_s^2 = 4.39 \times 10^{-6}$, so the noise robustness $R = \sigma_s^2 / \sigma_c^2 = 3.04$, which is close to the $R = 3$ predicted by the theory.
      
\begin{figure}[htb] 
	\includegraphics[width=1\linewidth]{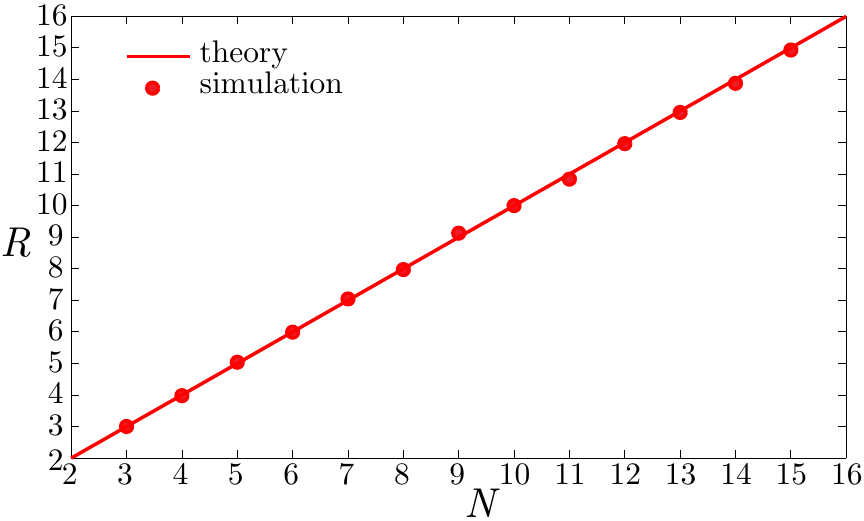}
	\caption{Six-step noise robustness $R$ versus lattice size $N$ of a GCM lattice of optimally coupled chaotic quadratic maps.} 
	\label{RGCMPlot}
\end{figure}

\begin{figure}[htb] 
	\includegraphics[width=1\linewidth]{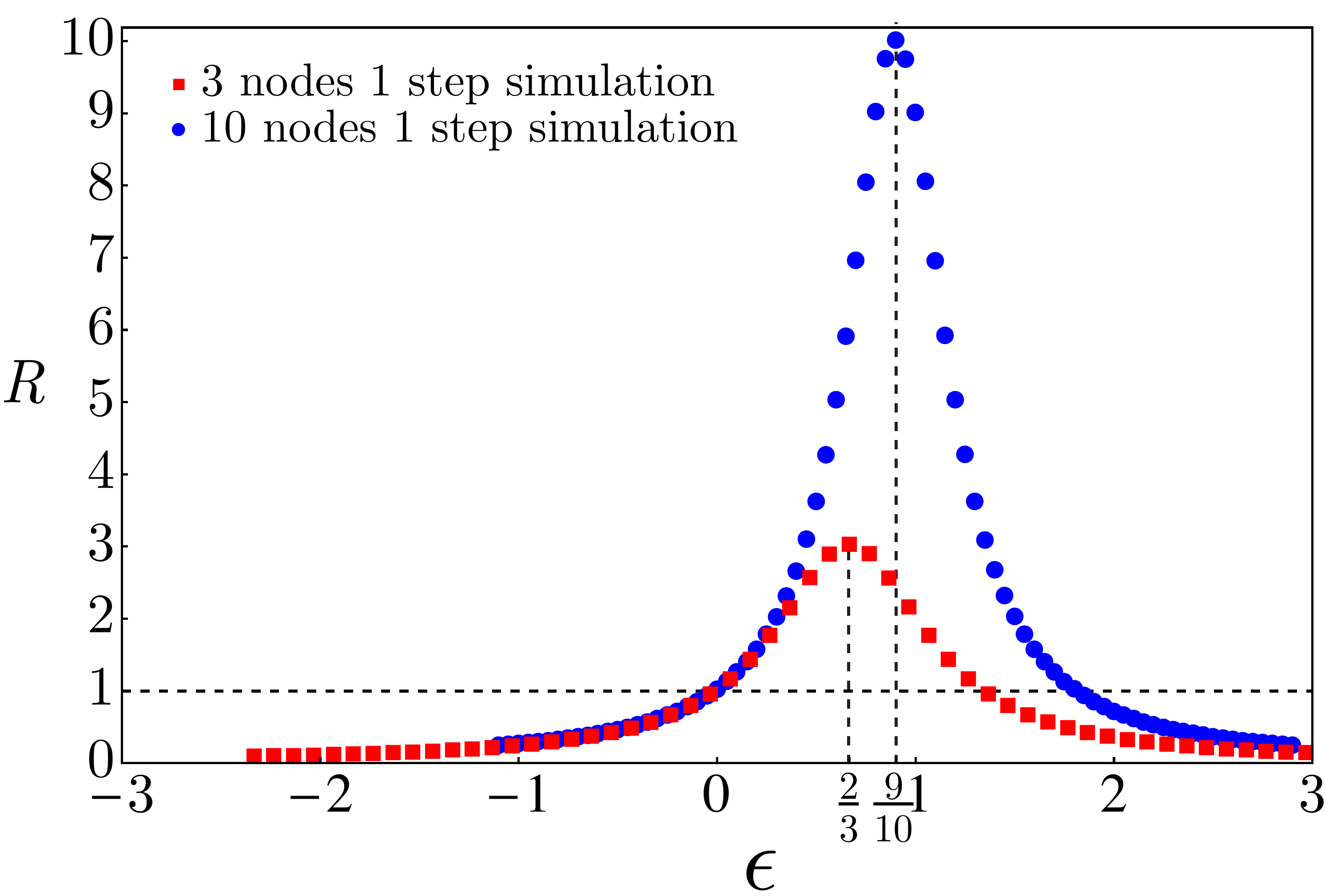}
	\caption{One-step noise robustness $R$ versus coupling parameter $\epsilon$ for lattice sizes of $N = 3$ and $N = 10$.} 
	\label{REpsilonPlot}
\end{figure}

To investigate the effect of lattice size $N$ on noise robustness $R$, we study a GCM lattice of $N$ Eq. (\ref{QuadMapEq}) quadratic maps with bifurcation parameter $\lambda = 2$ and coupling parameter $\epsilon = (N - 1) / N$. We initialize all $N$ maps of the GCM to a randomly chosen initial condition and then let the GCM iterate for a specific number of iterations and measure the deviation from noise free evolution. The additive noise variance in this example is $\sigma^2 = 10^{-6}$, and the iteration number is $i = 6$, but similar results can be obtained using other iteration numbers and noise variances. We repeat this process $10^6$ times, each time with a new random initial condition. At the end we measure the variances of noise deviations in the GCM for each lattice size, and calculate the ratios against the variance of noise deviations in a single isolated quadratic map. The results are depicted in Fig.~\ref{RGCMPlot}. We observe that in a GCM lattice of size $N$, noise robustness $R = N$, where $N=3, 4, \ldots,15$. While this is for the optimal coupling parameter, Fig.~\ref{REpsilonPlot} shows the noise robustness $R$ of the GCM lattices of sizes $N = 3$ and $N = 10$ for different coupling parameters when the GCM lattice iterates just once.
      
\begin{figure*}
	\includegraphics[width=1\linewidth]{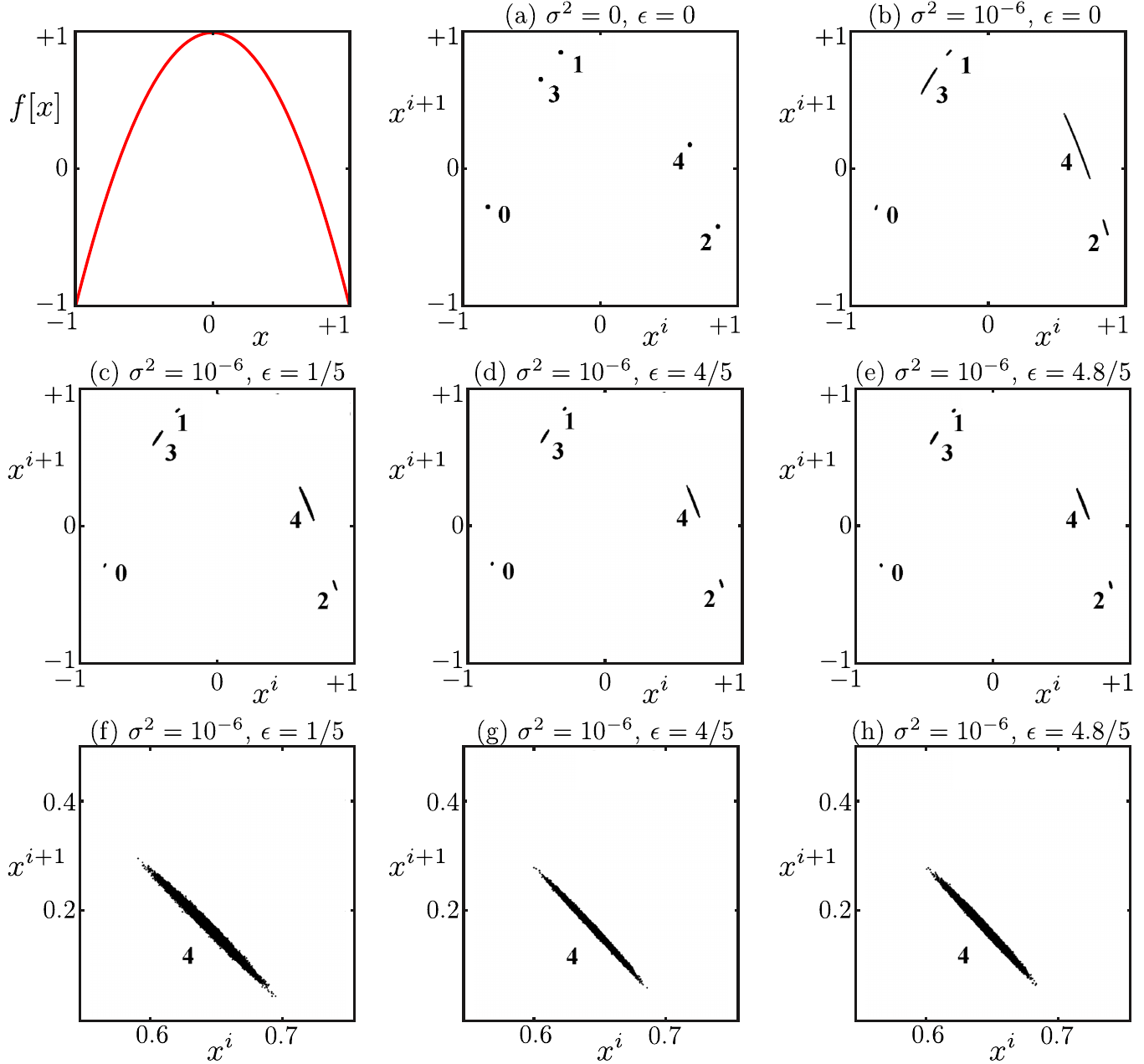}
	\caption{Delay coordinate embeddings of the first 5 steps of $10^5$ realizations of a GCM lattice of size $N = 5$ with the Eq.~(\ref{QuadMapEq}) quadratic map for different coupling parameters $\epsilon$. The optimal coupling of $\epsilon = 4/5$ produces the smallest final spread.} 
	\label{NoisyAttractorPlot}
\end{figure*}

We can also examine the effects of coupling by observing the noisy attractor of the coupled and uncoupled maps. For illustrative purposes, we consider a simple example. Figure~\ref{NoisyAttractorPlot} shows delay coordinate embeddings of the first 5 iterates of the Eq. (\ref{QuadMapEq}) quadratic map when $\lambda = 2$, where the dynamics is chaotic, from the initial condition $x^0 = -0.8$. Figure~\ref{NoisyAttractorPlot}(a) shows the iterates of an isolated single map (corresponding to $\epsilon = 0$) when noise variance $\sigma^2 = 0$. Figure~\ref{NoisyAttractorPlot}(b) shows $10^5$ different realizations of the iterates of the isolated single map (corresponding to $\epsilon = 0$) when the noise variance $\sigma^2 = 10^{-6}$. Figure~\ref{NoisyAttractorPlot}(c-e) shows the iterates of a map in a GCM of size $N = 5$ when the noise variance $\sigma^2 = 10^{-6}$ for coupling parameters $\epsilon = 1/5, 4/5, 4.8/5$. We observe that by coupling maps together the ``fuzziness" or noise content in the orbit is reduced. Furthermore, the maximum amount of noise reduction occurs when $\epsilon = 4/5$, which is the optimal coupling parameter based on the theory we developed in Section \ref{NoiseTheorySection}. For better comparison of the effects of the coupling value, Fig.~\ref{NoisyAttractorPlot}(f-h)  zooms in on the corresponding last iterate $x^4$ of Fig.~\ref{NoisyAttractorPlot}(c-e). The non-optimal coupling parameters produce thicker, noisier iterates.

So far we have demonstrated that for a GCM of size $N$, the coupled dynamics can reduces the variance of noise deviation to $1/N$ of variance of noise deviation in a single isolated map, producing a noise robustness of $R = N$. However there are exceptions to this rule. The major exception is when the dynamics is chaotic, and the evolution time is long enough so that the noise deviation becomes too large. For example, assume the case that the basin of attraction of a chaotic map is limited to one small portion of the state space, and beyond this portion the map diverges to infinity. In finite time, the noise can evolve to be large enough to push the orbits out of this basin and as a results these noisy orbits will diverge. The resulting ratio of variances of noise deviations could then be much different. We study this case here, where the dynamical system of Eq.~(\ref{QuadMapEq}) quadratic map has a small basin of attraction for its chaotic attractor, $[-1, 1]$, but beyond this interval the map diverges to infinity. (In Section~\ref{SineMapSection}, another dynamical system will be studied where the entire state space is the basin of attraction for the chaotic attractor, and there another exception to our rule will be studied.) When a noisy orbit leaves this basin of attraction, it diverges from the chaotic attractor, usually at an exponential rate. As a result, one can argue that the noisy orbits of such systems have two phases: first, when they are still inside the basin of attraction of the chaotic attractor; second, when they have eventually pushed out of the basin of attraction because of noise. 

For example, consider Eq.~(\ref{QuadMapEq}) quadratic map with initial condition $x^0 = 0.7$ and variance of the additive noise $\sigma^2 = 10^{-6}$. We calculate the noise robustness $R$ in a CML of size $N = 3$ for different iteration numbers $i$. The results are depicted in Fig. 6. We observe that for the first few iterations, the ratio is 3 as we expected. But from iteration 7 onwards the iterates of the single map start to fall beyond the basin of attraction and escape, whereas the iterates of the CML still remain on the attractor, basically because the coupled dynamics averages the local noise terms and reduces its effects. The ratio of variances of the noise deviations exponentially increases. However upon further iteration the noise effects in CML will become large enough to push the orbits beyond the basin of attraction as well. Numerical simulation, with finite precision, doesnÕt return any valid number for the ratio of variances for these cases because the variances are too large.
      
\begin{figure}[htb] 
	\includegraphics[width=1\linewidth]{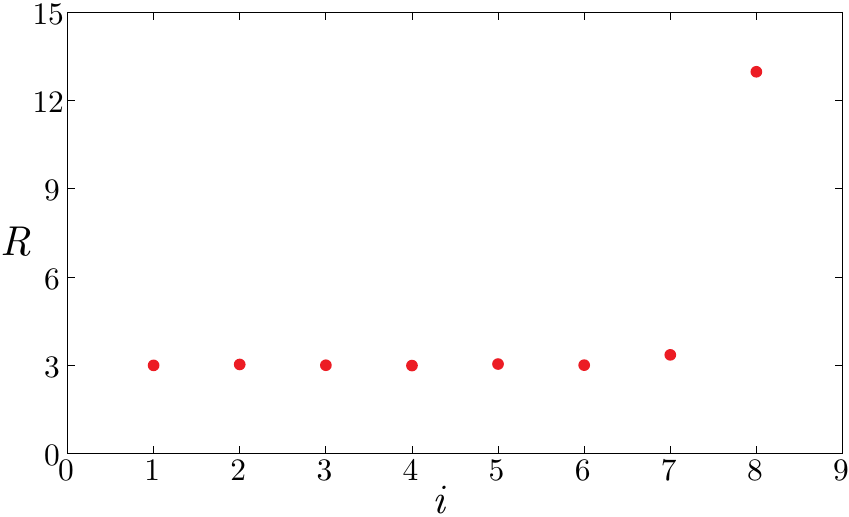}
	\caption{Noise robustness $R$ versus iteration number $i$ for $N = 3$ coupled chaotic quadratic maps. By iteration $i = 8$, the single noisy quadratic map has escaped its initial basin of attraction and has begun diverging.}
	\label{REscapePlot}
\end{figure}

\begin{figure}[htb] 
	\includegraphics[width=1\linewidth]{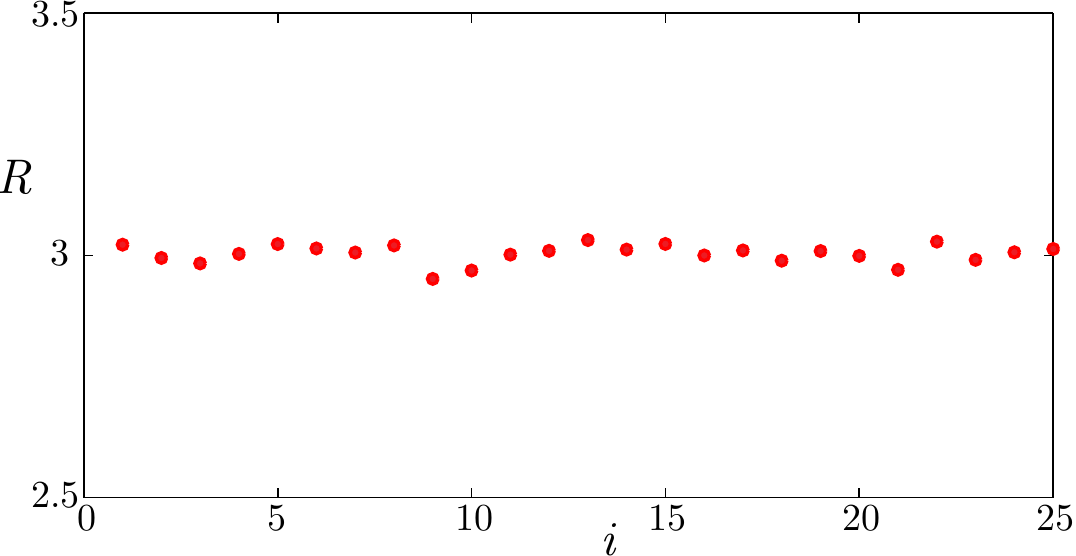}
	\caption{Noise robustness $R$ versus iteration number $i$ for $N = 3$ coupled \textit{periodic} quadratic maps.}
	\label{RIPlot}
\end{figure}

The observed limitations discussed above are \textit{not} due to the coupled dynamics but simply due to the nature of the averaging filter and chaotic dynamics. Even with the application of a conventional averaging filter to average the dynamical states of a set of independent maps (at each iteration) and then using this resulting value to initialize the maps to perform a new iteration, there is still noise remaining in the averaged states. This noise evolves exponentially over time, and eventually leads the iterates of the map out of the basin of attraction of the chaotic attractors. Alternately, exponentially growing noise can cover the entire chaotic attractor, as we will demonstrate for the coupled dynamics based realization of the averaging filter in Section~\ref{SineMapSection}. 

If the dynamics is not chaotic, the noise effect doesnÕt grow exponentially over time to rapidly push the orbit out of the basin of the attractor. As a result the expected noise robustness can be obtained for higher iterates of the CML as well. To examine this case, we have studied a CML of size $N = 3$, when the local dynamics is governed by the Eq.~(\ref{QuadMapEq}) quadratic map but with bifurcation parameter $\lambda = 1.2$ where the dynamics is periodic. In Fig.~\ref{RIPlot} shows the noise robustness $R$ for different iteration numbers $i$ when the variance of the additive noise is $\sigma^2 = 10^{-6}$. The noise robustness $R = 3$ even for higher iteration numbers of the CML.

\subsection{Sine Map} \label{SineMapSection}

To demonstrate that noise reduction in CMLs is not limited to a specific type of local dynamics, we next study the sine map
\begin{equation} \label{SineMapEq}
	x^{i + 1} = r \sin \left[ \pi x^i \right],
\end{equation}
which is chaotic on the interval $x^i \in [0,1]$ for bifurcation parameter $r = 1$. The key difference between the Eq.~\ref{QuadMapEq} quadratic map and the Eq.~\ref{SineMapEq} sine map is that the quadratic map has a finite basin of attraction and the sine map has an infinite basin of attraction, and this affects the evolution of the noise robustness $R$. 

\begin{figure}[htb] 
	\includegraphics[width=1\linewidth]{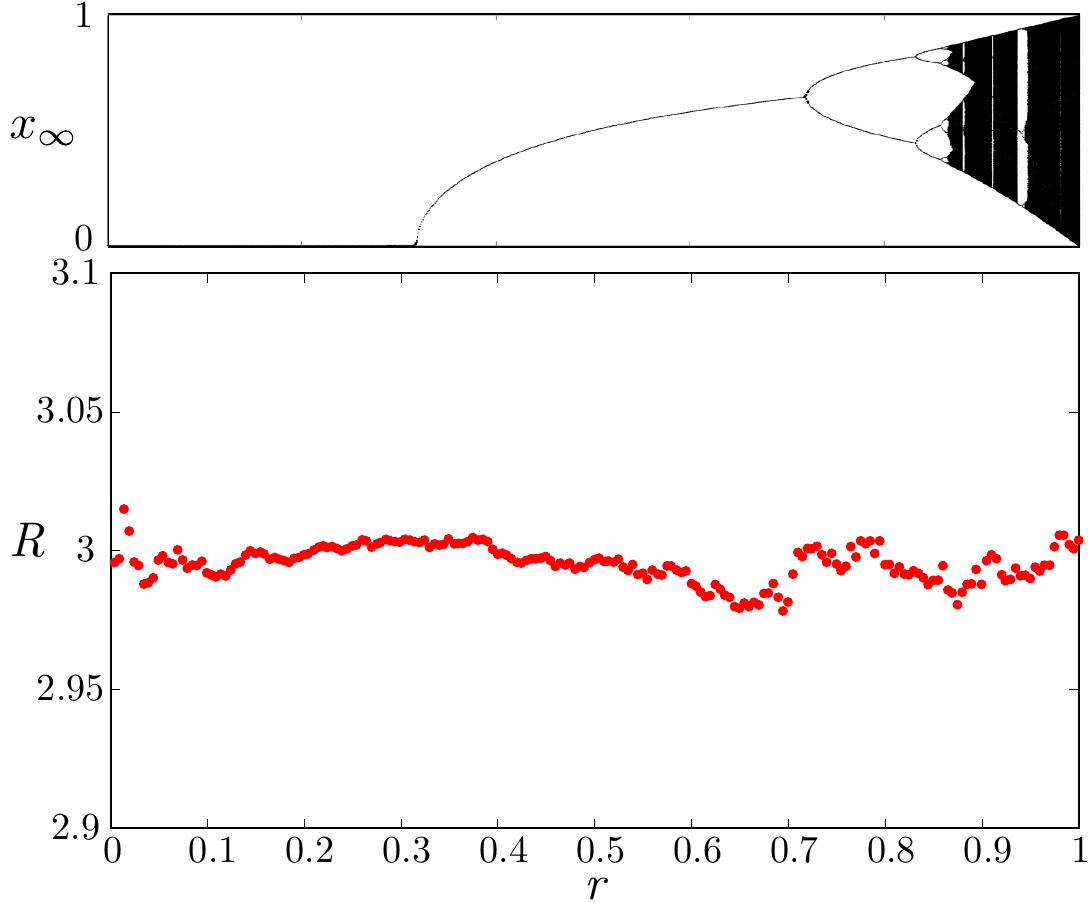}
	\caption{Five-step noise robustness $R$ versus bifurcation parameter $r$ for $N = 3$ coupled sine maps (bottom); sine map bifurcation diagram (top).}
	\label{RISinePlot}
\end{figure}

The sine map exhibits different types of dynamics depending on the bifurcation parameter $r$. Figure~\ref{RISinePlot} presents the corresponding bifurcation diagram. To demonstrate that the reported phenomenon of noise reduction in CMLs is not just restricted to a specific regime of dynamics, but rather generic, we statistically compute the noise robustness $R$ of coupled maps for different values of bifurcation parameter $r$. That result is presented in Fig.~\ref{RISinePlot}. We observe that no matter what the regime of the local dynamics, the noise robustness $R \approx 3$. In this simulation, we randomly initialize all $N = 3$ nodes of the lattice to a common point $x^0 \in [0,1]$. From this random initial condition we evolve the coupled map lattice, and after $i = 5$ iterations we stop the evolution and measure how much the final state of a map of the CML has deviated from noise free evolution. We repeat this process $10^6$ times, each time with a new random initial condition, and calculate the variance of these noise deviations. Then we repeat the same procedure again, but this time with a single isolated map, and compute the variances of the noise deviations. The ratio is the noise robustness $R$.

\begin{figure}[htb] 
	\includegraphics[width=1\linewidth]{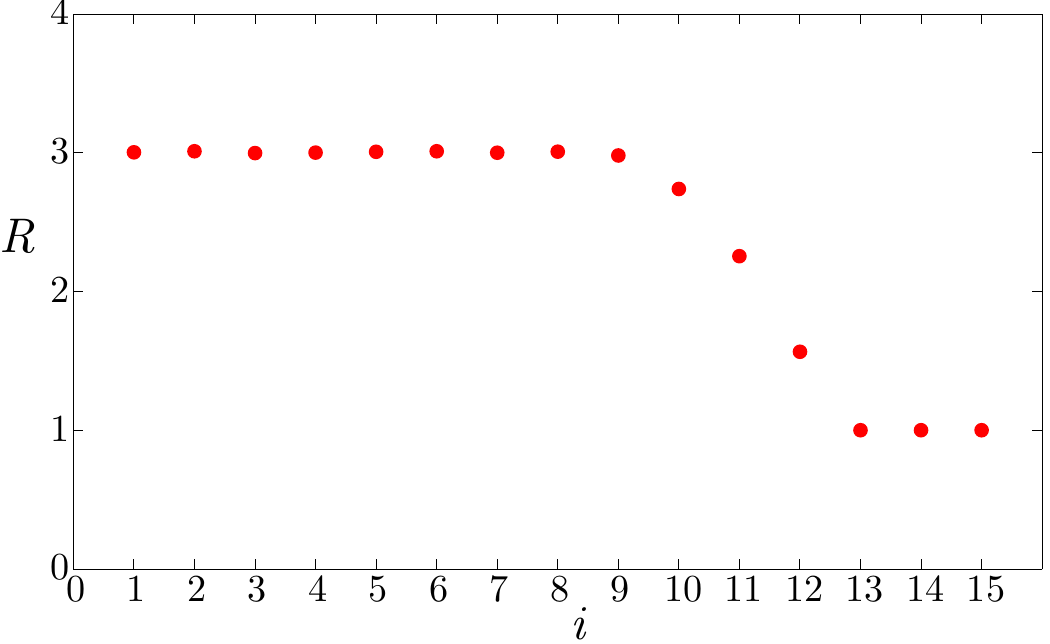}
	\caption{Noise robustness $R$ versus iteration number $i$ for $N = 3$ coupled sine maps. By iteration $i =13$, noise has diffused over the entire attractor.}
	\label{RISineDropPlot}
\end{figure}

For bifurcation parameter $r = 1$, the Eq.~(\ref{SineMapEq}) sine map has a chaotic attractor whose basin of attraction is the entire state space. Therefore there is no escaping orbit from the chaotic attractor and the condition depicted in Fig.~\ref{REscapePlot} doesnÕt happen. But a new exception to the Section~\ref{NoiseTheorySection} theory happens in this chaotic system. In a chaotic system, in finite time, noise effects will spread over the entire chaotic attractor. This condition happens earlier to a single map and later to the CML. Figure~\ref{RISineDropPlot} depicts the noise robustness $R$ for a CML of size $N = 3$. Initially, noise robustness $R = 3$, as we expect. But after about $i = 9$ iterations, the noise in a single map has already spread over the entire attractor, and therefore its variance doesnÕt increase anymore. But the noise in the CML has not covered the attractor yet and its variance is still growing with each iteration. As a result the ratio of variances decreases. It takes $i = 13$ iterations for noise in CML to cover the entire attractor.  Now the variances of noise effects for both single map and the CML are exactly the same: the size of the attractor. Therefore, from iteration $i  = 13$ and thereafter the noise robustness is $R = 1$ and remains unity thereafter.

\begin{figure}[htb] 
	\includegraphics[width=1\linewidth]{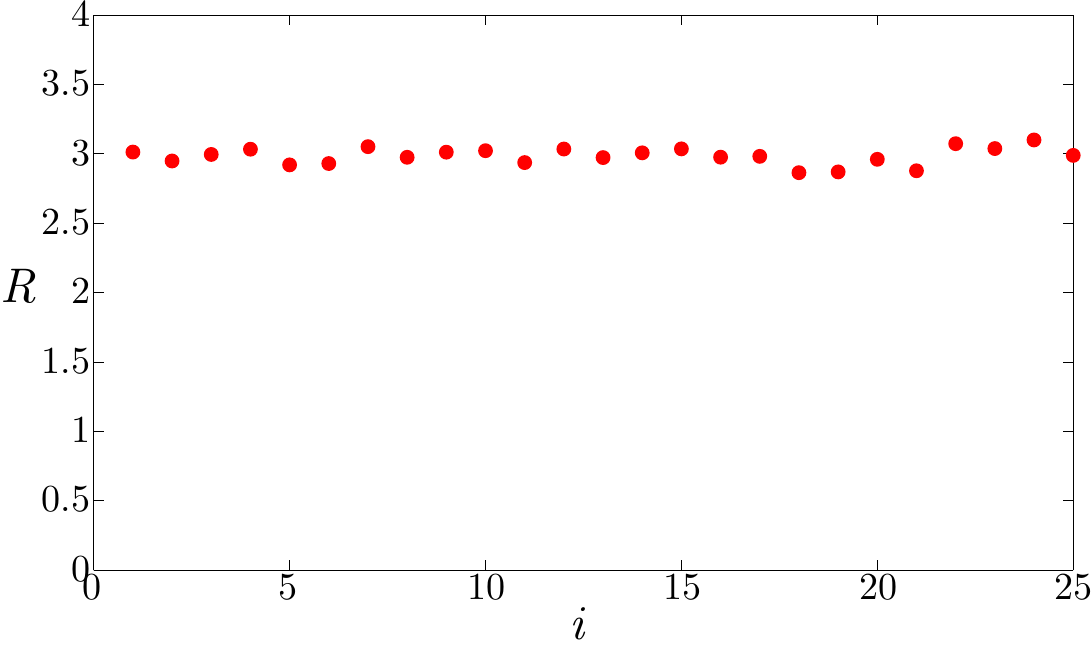}
	\caption{Noise robustness $R$ versus iteration number $i$ for $N = 3$ coupled \textit{periodic} sine maps. The robustness is independent of the iteration number.}
	\label{RISinePeriodicPlot}
\end{figure}

When the dynamics is not chaotic, the noise doesnÕt grow exponentially over time to rapidly cover the entire attractor (Fig. 9) or to push the orbit out of the basin of the attractor (Fig. 6). Therefore we can iterate the single map and the CML longer and still get the expected noise robustness $R$.  To investigate this case, we have studied a CML of size $N = 3$ the Eq.~(\ref{SineMapEq}) sine map at  bifurcation parameter $r= 0.8$, where the dynamics is periodic. Figure~\ref{RISinePeriodicPlot} shows the noise robustness $R$ for different iteration numbers  $i$ when the variance of the additive noise is $10^{-6}$. We observe that the noise robustness $R \approx 3$ even for large iteration number of the CML.

\subsection{Local Coupling Noise Diffusion}

\begin{figure}[htb] 
	\includegraphics[width=1\linewidth]{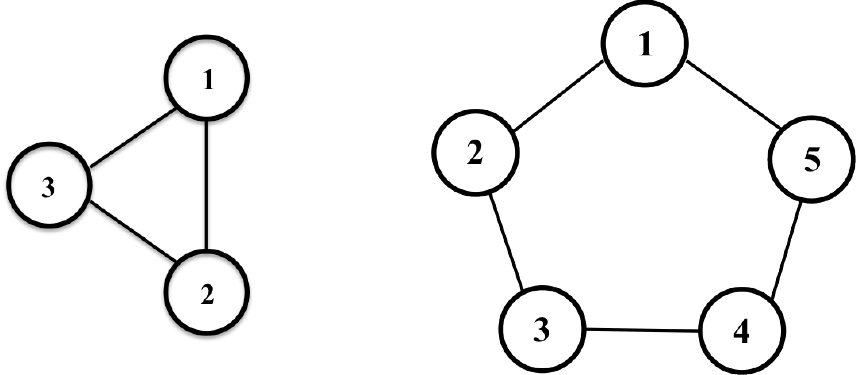}
	\caption{Schematic three-node CML with local \textit{and} global connectivity (left) and five-node CML with local, nearest-neighbor coupling (right). Each node (circle) is coupled (lines) to its two nearest neighbors.}
	\label{FiveNodePlot}
\end{figure}

Global coupling is not necessary to reduce the noise effects in a CML. Noise can diffuse across the lattice through local coupling as well, and eventually the effects of local noise from different nodes attenuate and reduce each other and the coupled dynamics can still function as an averaging filter. The noise evolution and noise robustness of a CML of size $N = 5$ with local coupling is studied in this section. Figure~\ref{FiveNodePlot} depicts the architecture of the CML with local connectivity. The coupling scheme is
\begin{equation} \label{LocalCouplingEq}
	x_n^{i+1} = \frac{1}{3} f\left[ x_n^i \right] + \frac{1}{3} \left( f \left[ x_{n-1}^i \right] +  f \left[ x_{n+1}^i \right] \right).
\end{equation}

\begin{figure}[htb] 
	\includegraphics[width=1\linewidth]{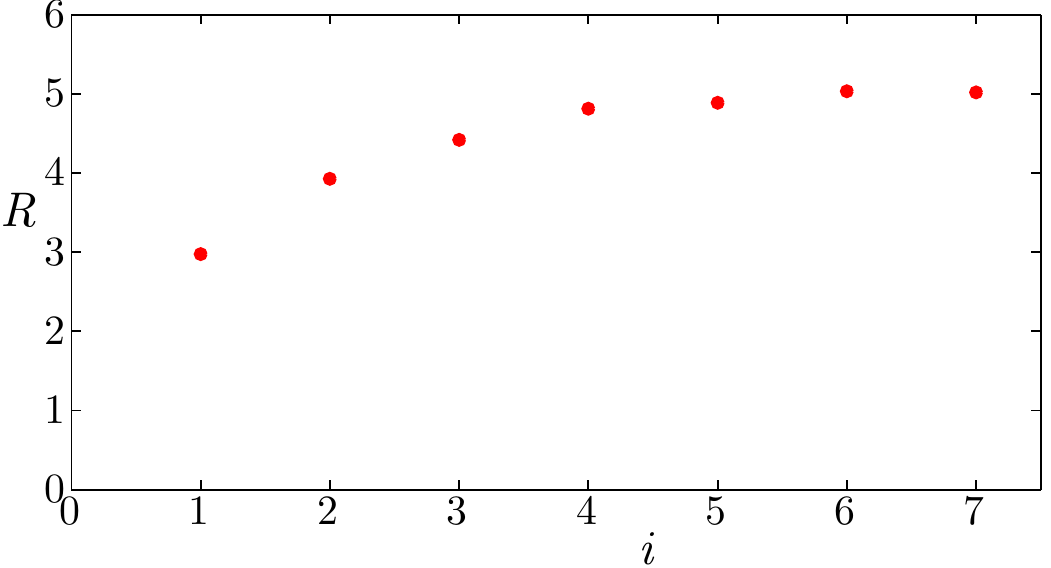}
	\caption{Noise robustness $R$ versus iteration number $i$ for $N = 5$ nearest-neighbor coupled quadratic maps. By iteration $i = 5$, noise robustness peaks at $R = 5$, even though the coupling averages only the nearest neighbors.}
	\label{NIQuadLocalPlot}
\end{figure}

As a first example, we let the Eq.~\ref{QuadMapEq} quadratic map with bifurcation parameter $\lambda = 2$ and additive noise variance $\sigma^2 = 10^{-6}$ determine the local dynamics. Figure~\ref{NIQuadLocalPlot} depicts Monte Carlo simulation results for noise robustness. At the first iteration, we observe that the noise robustness is three, which is expected. The coupling is local, and each node is connected to its left and right neighbors. Therefore the noise effects are averaged over three nodes. But as the CML iterates, the noise effects diffuse across the lattice, and eventually the noise robustness peaks at $R = 5$, which is the size of the lattice. This means that global coupling is not necessary for global diffusion of noise, rather the noise can diffuse though local coupling, and eventually local noise effects from different nodes average each otherÕs effects and the result is a system with less noise content. 

After about $i = 8$ iterations the noise starts to cover the entire attractor and then pushes the orbits beyond the basin of attraction. This is the same phenomenon that we observed and reported for GCM lattice in Section~\ref{QuadMapSubSection}. 

\begin{figure}[htb] 
	\includegraphics[width=1\linewidth]{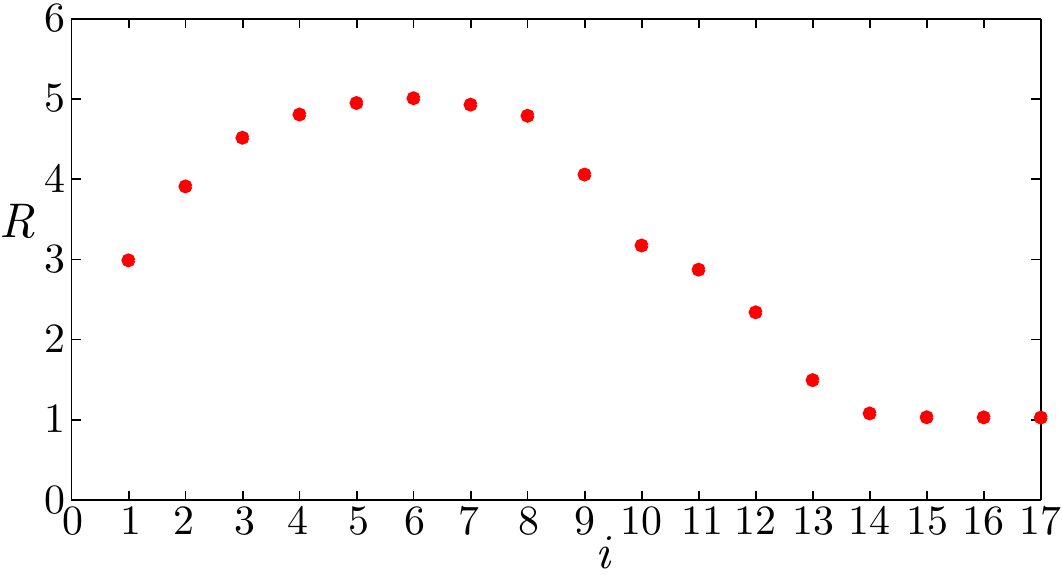}
	\caption{Noise robustness $R$ versus iteration number $i$ for $N = 5$ nearest-neighbor coupled sine maps. By iteration $i = 5$, noise robustness peaks at $R = 5$, even though the coupling averages only the nearest neighbors. By iteration $i=14$, noise has diffused over the entire attractor.}
	\label{NISineLocalPlot}
\end{figure}

As a second example, we let the Eq.~\ref{SineMapEq} sine map with bifurcation parameter $r = 1$ and additive noise variance $\sigma^2 = 10^{-6}$ determine the local dynamics. Figure~\ref{NISineLocalPlot} depicts Monte Carlo simulation results for noise robustness.

Similar to Fig.~\ref{NIQuadLocalPlot}, at the first iteration, we observe that the noise robustness is $R = 3$. The coupling is local, and therefore the noise effects are averaged over three nodes. But as the CML iterates, the noise effects diffuse across the lattice through the local coupling, and as a result, eventually the noise robustness peaks at $R = 5$, which is the size of the lattice. This again implies that local coupling is enough for diffusion of noise and building a more robust to noise system.

Similar to the noise robustness of a CML of size $N = 3$ shown in Fig.~\ref{RISineDropPlot}, at iteration $i = 8$ the noise robustness $R$ decreases and eventually converges to one. The reason behind this observation in a locally coupled map lattice is exactly the same as discussed in Section~\ref{SineMapSection}. The noise grows and eventually covers the entire attractor in both single map and the coupled map. Therefore the ratio of noise deviations, which is the noise robustness $R$, becomes unity.  
      
\section{Conclusions} \label{Conclusions}

In this paper, we have demonstrated how coupling nonlinear dynamical systems can reduce the effects of noise. For Kaneko three-node coupled map lattices, we showed in theory and in simulations how to tune the coupling to optimize the noise reduction and create a dynamics-based averaging filter. For larger lattices, we generalized the noise reduction to both global and nearest-neighbor coupling. We also explored factors that modulate the noise reduction, including the periodicity or chaoticity of the underlying maps, the sizes of their basins of attraction, and their transient and steady states.

Coupled dynamics can realize an averaging filter that can be used in dynamics-based applications such as chaos computing, chaos communications, and chaos based optimizers to design more noise tolerant versions of these applications. The coupling obviates the need for a dedicated averager to reduce the noise content. Instead, the inherent coupled dynamics achieves the same effect. In chaos computing applications, the noise reduction thereby becomes part of the chaotic computing architecture itself rather than an addition to it. 

For computational speed and theoretical simplicity, this paper focussed on coupled map lattices, but the phenomenon of coupling reducing noise is more general and applies to many other kinds of systems. Having the ability to exploit coupled dynamics for noise reduction offers the opportunity to develop and design dynamical systems that can perform dynamics-based application while reducing unwanted noise through the dynamics itself.
      
\begin{acknowledgments}
We gratefully acknowledge support from the Office of Naval Research under Grant No. N000141-21-0026 and STTR grant No. N00014-14-C-0033.
\end{acknowledgments}



\end{document}